 \documentclass[review,1p,times]{elsarticle}

\usepackage{graphicx}
\usepackage{amssymb}
\usepackage{xcolor}
\usepackage{xurl}
\usepackage{caption}
\usepackage{dirtytalk}
\usepackage{array}
\newcolumntype{P}[1]{>{\centering\arraybackslash}p{#1}}
\usepackage{float}
\usepackage{lineno} 

\journal{Continental Shelf Research}

\begin{document}

\begin{frontmatter}


\title{Modeling the seasonal variability and the governing factors of Ocean Acidification over the Bay of Bengal region}



\author{ A.P Joshi\corref{label1}\textsuperscript{1*} and H.V Warrior\corref{cor2}\textsuperscript{1}}

\address{\textsuperscript{1}Department of Ocean Engineering and Naval Architecture, IIT Kharagpur, Kharagpur-721302, West Bengal, India}

\cortext[label1]{apurvajoshi@iitkgp.ac.in}

\begin{abstract}
The Bay of Bengal (BoB) is a high recipient of freshwater flux from rivers and precipitation, making the region strongly stratified. The strong stratification results in a thick barrier layer formation, which inhibits vertical mixing making this region a low-productive zone. In the present study, we attempt to model the pH of the BoB region and understand the role of different governing factors such as sea-surface temperature (SST), sea-surface salinity (SSS), dissolved inorganic carbon (DIC), and total alkalinity (TALK) on the seasonality of sea-surface pH. We run a set of sensitivity experiments to understand the role of each of the governing factors. The results show that the SST, SSS, and DIC are the principal drivers affecting the sea-surface pH, while TALK plays a buffering role. The SST and DIC are consistently found to be opposite to each other. The pre-monsoon season (MAM) has shown to have an almost equal contribution from all the drivers. In the pre-monsoon season, the SST and DIC are balanced by TALK and SSS. The role of SSS is significantly dominant in the second half of the year. Both SST and SSS counter the role of DIC in the southwest monsoon season. The strong stratification plays an essential role in modulating the pH of the BoB region. The thickness of the barrier layer formed in the sub-surface layers positively affects the sea-surface pH. The northern BoB is found to be more alkaline than the southern BoB. Our study highlights the complexity of ocean acidification in the BoB region compared to the other part of the world ocean.
       
\end{abstract}


\end{frontmatter}


\section{Introduction}
\label{S:1}

The pronounced absorption of the anthropogenic atmospheric carbon dioxide by oceans (\ensuremath{2.5 \pm 0.6} GtC yr\textsuperscript{-1}, \citep{friedlingstein2020global}) causes a drop in oceanic pH referred to as Ocean Acidification. A few studies indicate a drop of 0.1 units in the upper ocean pH, and it is predicted to drop by almost 0.3 to 0.5 units by the end of this century \citep{sabine2004oceanic,feely2009ocean,kwiatkowski2018diverging}. This reduction in sea-surface pH has a major effect on the biological process and production in oceans \citep{gattuso2014ocean}.These long-term variations in oceanic pH values include short-term fluctuations (diurnal, seasonal, and interannual). The extent of vulnerability of these short-term fluctuations is equal or sometimes greater than the long-term changes \citep{provoost2010seasonal,wootton2012carbon,sutton2014high}. Hence to improve the predictions of ocean pH variability, identification of drivers affecting it is of utmost importance.

The Bay of Bengal (BoB) region chosen in this study is peculiar by reference to its geographical settings (enveloped by land from the three sides and open in the south), the substantial freshwater influx from rivers and precipitation \citep{unesco1969discharge}, and the seasonal reversal of coastal currents \citep{murty1993current,shetye1996hydrography}. The abundance of freshwater in the BoB region makes this region one of the challenging zones for ocean acidification studies. The ocean acidification scenario for the BoB region, as reported by \cite{feely2009ocean}, suggests the pH to be below 8.0 by 2050 and may go below 7.8 in 2095. Since the BoB is a reservoir of an abundance of marine species, especially shells and coral reefs, the ocean acidification scenario presented by \cite{feely2009ocean} is of great worry to the scientist, environmentalists, and policymakers. 

\cite{mukhopadhyay2002seasonal} noted the lowest pH in Mahanadi estuaries (\ensuremath{7.02 \pm 0.1}) and highest in the Ganges estuaries (\ensuremath{8.13 \pm 0.24}). \cite{sarma2012sources}, based on ship observations in the western coast of BoB, showed that the northwestern coast has higher pH (\ensuremath{8.45 \pm 0.003}) than the southwestern coast (\ensuremath{8.12 \pm 0.03}). The lowest pH was observed in the Godavari estuaries (\ensuremath{8.07 \pm 0.05}), while the Mahanadi estuaries demonstrated the highest pH values (\ensuremath{8.41 \pm 0.005}).  The sharp gradient between the northwestern and southwestern pH values are associated with the difference in salinity (\ensuremath{31.96 \pm 0.88} for Godavari estuaries and \ensuremath{23.24 \pm 1.84} for Mahanadi estuaries).  

\cite{sarma2015observed} compares the measured hydrographic properties and inorganic carbon components of the BoB between March-April of 1991 and 2011, this work also includes a time series observation of 8 years from 2005 to 2013. All these observations were of the western coast of BoB. The study suggests that the rate of decrease in pH in the southwestern coast was consistent with the global trend, but the northwestern coast experienced a higher decrement rate of pH (3 to 5 times higher).  The time series observation at Vishakhapatnam (located on the southwestern coast of BoB) and Paradip (located on the northwestern coast of BoB) demonstrates a decline in the pH at -0.0015 units year\textsuperscript{-1} and -0.005 units year\textsuperscript{-1}, respectively. The higher declining rate in at the Paradip station is attributed to enhanced aerosol depositions in the BoB.

The seasonal reversing coastal currents are known as the East India Coastal Currents (EICC) or Western Boundary Currents (WBC), play a significant role in controlling the sea-surface salinity (SSS) of the BoB region. From February to May, the northward moving EICC increases the salinity, which weakens the stratification, and enhances coastal upwelling, resulting in a decrease of sea-surface pH \citep{sarma2018east}. The EICC flows southwards during October to December, which brings in low saline, more basic waters from north to the western coast of BoB, that increases the pH of this zone \citep{sarma2018east}.  Since the circulation pattern of BoB consists of many eddies, they play a significant role in modulating the sea-surface pH. \cite{sarma2019impact}, based on the observation in the western BoB, reports the presence of low pH values in the cyclonic eddy and no-eddy regions (upper 200m) of BoB, whereas the high pH waters extends up to 150 m to 175 m deep in the anticyclonic eddy regions.

Using Multiple Linear Regression (MLR) \cite{sridevi2021role} shows an increase in pH in the BoB (1998-2015), the only exception being the head bay region. The near proximity of the head bay to the land makes it vulnerable to the atmospheric and river pollutants. The acidification of the head bay is associated with the rise in sea-surface temperature and atmospheric depositions, as well as decrease in freshwater discharge. The SSS plays a major role in controlling the seasonality of the northern BoB (seasonal amplitude of $\approx$ 0.18 units) \citep{chakraborty2021seasonal}. \cite{chakraborty2021seasonal} using interannual model for the North Indian Ocean (NIO) shows the contribution of the various drivers on the seasonality of total pH and \textit{p}CO\textsubscript2 for the Arabian as well as the BoB region.  

In this study, we aim to model the seasonality of pH (based on the climatological modeling) for the whole Bay of Bengal region and explore the physical factors driving the seasonality of pH in this region. As the BoB is recipient of large freshwater influx from precipitation and rivers, it affects the stratification dynamics of the BoB. Through this study we attempt to analyze the role of different drivers influencing the pH variability and the role of stratification in monitoring the pH variability. The remaining paper is structured as follows: Section 2 describes the Data and Methodology used in this study; Results and Discussion are presented in section 3; Conclusion in section 4. 

\section{Data and Methodology}
\label{s:2}

\subsection{Model}
\label{s:2.1}

In our previous study \citep{joshi2020configuration}, we exhibit a comparison between two types of model configurations (the first one using the \cite{fairall1996bulk} bulk formulation to calculate the wind stress and evaporation. In the second configuration, we externally provide the Wind Stress and evaporation minus precipitation (E-P) data). We concluded that the method of externally providing the wind stress and E-P data emulated the carbonate chemistry satisfactorily. We provide an exhaustive evaluation of the different physical (current, sea-surface temperature, sea-surface salinity, barrier layer thickness, mixed layer depth) and carbonate parameters (dissolved inorganic carbon, total alkalinity, pH, and \textit{p}CO\textsubscript2) in \cite{joshi2020configuration}. Using the same model, we further explore the influence of the freshwater plume spreading and the barrier layer thickness on the sea-surface \textit{p}CO\textsubscript2 \citep{joshi2021influence}. Using this model we also explore the different mechanisms and drivers affecting the BoB \citep{joshi2022comprehending}. Since this study is an extension of our previous work, to avoid recapitulation we provide a summarized model configuration information. We encourage readers to refer to our previous works \citep{joshi2020configuration,joshi2021influence,joshi2022comprehending} for a comprehensive model information. 

\begin{figure}[ht!]
\centering
\includegraphics[height=11cm,width=10cm]{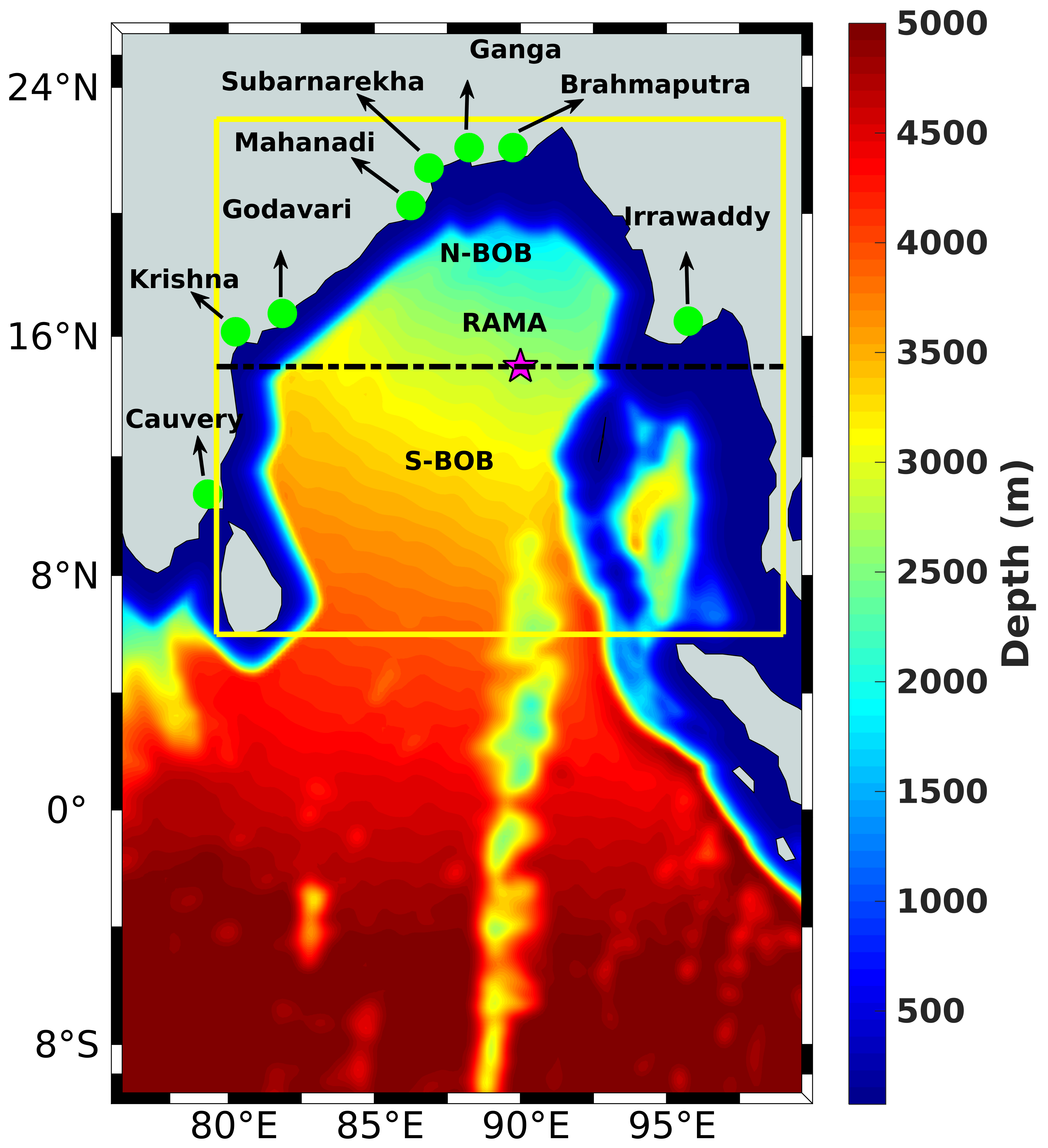}
\caption{ The contours represent the model bathymetry. The yellow box marks the study region, while the red star shows the Rama buoy location. The broken black line divides the study region into the Northern Bay of Bengal (N-BoB) and the Southern Bay of Bengal (S-BoB). The rivers included in this study have been demonstrated using green dots.}
\label{fig:1}
\end{figure}
 
 The present study couples the Regional Oceanic Modeling System (ROMS) (physical model) \citep{shchepetkin2005regional,shchepetkin2009correction} to the Pelagic Interaction Scheme for Carbon and Ecosystem Studies (PISCES) (biogeochemical model) \citep{aumont2003ecosystem,aumont2006globalizing}. Fig.\ref{fig:1} reveals the study region, which is considered in this paper. The contours show the bathymetry (extracted from ETOPO2 \citep{smith1997global}), and the yellow box indicates the domain of analysis. The rivers with high discharge volume are only included in this study (refer to Fig.\ref{fig:1}). The climatological river discharge data is included in the coupled model from \cite{dai2009changes,dai2013some}. The horizontal resolution is \ensuremath{1/7^o} with 32 sigma layers as vertical resolution.   
 
 Near the surface, refinement is achieved by assigning 12 vertical layers in the top 100 m at the maximum depth position. The K-profile parameterization turbulence scheme (KPP model) \citep{large1994oceanic} is adopted in the present study. We choose the physical tracers for initial and lateral forcing from the World Ocean Atlas, 2009 (WOA09) \citep{antonov2010world,he2010world}. The Comprehensive Ocean-Atmospheric Data Set (COADS) \citep{worley2005icoads} is exploited to provide the heat fluxes, temperature, humidity (both relative and specific), and density of air. The E-P and shortwave radiations are also provided from the COADS data set. The monthly climatology of winds and wind stresses are used from the QuikSCAT satellite scatterometer from 1999-2009 \citep{liu1998nasa,risien2008global}.
 
 We provide the nutrients and oxygen for the upper ocean from the WOA09 dataset, while for the interior oceans, oxygen and nutrients from the World Ocean Atlas PISCES (WOAPisces) \citep{aumont2003ecosystem} are used. The carbonate parameters (Dissolved Inorganic Carbon (DIC), Total Alkalinity (TALK), Dissolved Organic Carbon (DOC), etc.) are taken from the WOAPisces dataset. The recent parameterization scheme of \cite{echevin2008seasonal} is used for the PISCES model.
 
 \begin{figure}[ht!]
\centering
\includegraphics[height=9cm,width=\textwidth]{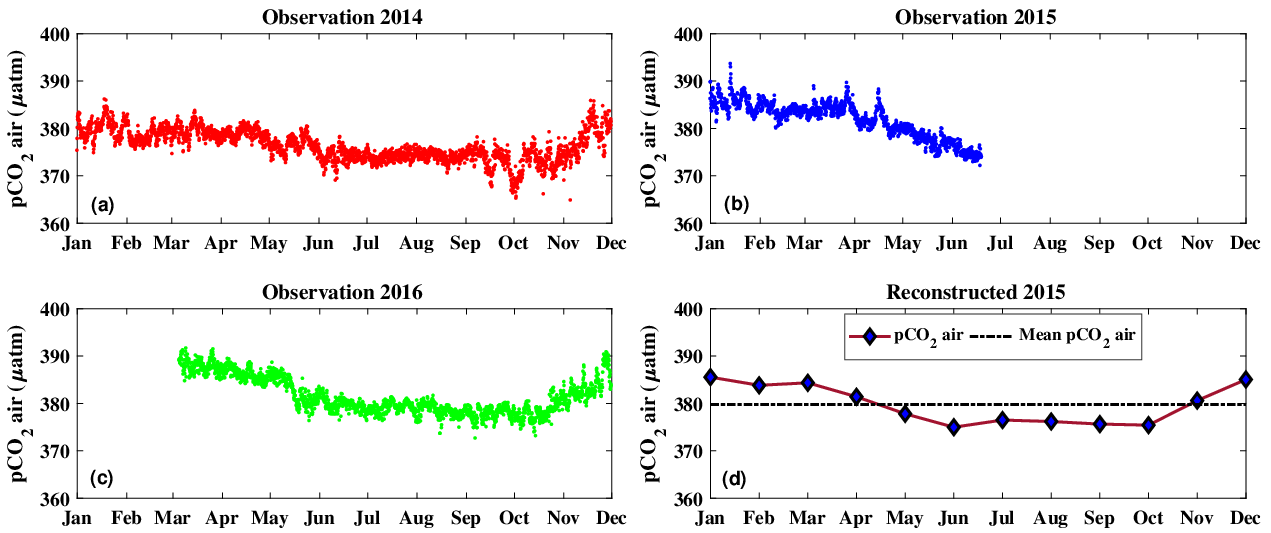}
\caption{The atmospheric \textit{p}CO\textsubscript2 from the Rama buoy for the year 2014 (panel a), 2015 (panel b), and 2016 (panel c). The reconstructed atmospheric \textit{p}CO\textsubscript2 of the year 2015 is provided in panel d. The black broken line shows the mean atmospheric \textit{p}CO\textsubscript2.}
\label{fig:2}
\end{figure}

For the atmospheric \textit{p}CO\textsubscript2, we analyze the open ocean Rama buoy values (Fig.\ref{fig:2}). As the observations are available for the 2014 full year, whereas January-June for 2015 and March-December for 2016, we reconstruct the atmospheric \textit{p}CO\textsubscript2 for 2015 using the available data. The reconstructed 2015 atmospheric \textit{p}CO\textsubscript2 data contains the original values for the January to June period, while the second half of the year is constructed by averaging the observations from 2014 and 2016. Fig.\ref{fig:2} d shows that the fluctuations in the atmospheric \textit{p}CO\textsubscript2 are relatively small, and the mean is 379.79 \ensuremath{\mu atm}. Since the previous study \citep{joshi2020configuration} shows that the carbonate parameters are reasonably emulated by the coupled model using 377 \ensuremath{\mu atm} (\ensuremath{< 1 \%} from the mean atmospheric \textit{p}CO\textsubscript2 of the reconstructed 2015 year), we continue using 377 \ensuremath{\mu atm} for the present study.

We run the ROMS model for 30 years before adding the PISCES model. Then both the models coupled together are further run for another 30 years. Finally, we average the last three years to create the climatology of the BoB region.

\subsection{pH calculation and sensitivity of pH seasonality to its governing factors}
\label{s:2.2}

The sea-surface pH can be factorized as the effect of SST, SSS, DIC, and TALK \citep{takahashi2009climatological,valsala2015mesoscale,sreeush2019variability}. The effect of nutrients and other minor ions (like borate, sulfate, and fluoride) on the sea-surface pH can be neglected and treated as residuals \citep{hagens2016attributing}. Hence the sea-surface pH in its linearized form can be written as:

\begin{equation}
    \frac{dpH}{dt} = \frac{\partial pH}{\partial DIC} \frac{dDIC}{dt} + \frac{\partial pH}{\partial TALK} \frac{dTALK}{dt} + \frac{\partial pH}{\partial SST} \frac{dSST}{dt} + \frac{\partial pH}{\partial SSS} \frac{dSSS}{dt}
\end{equation}
\label{e:1}

Based on the above equation, we construct the sea-surface pH for the BoB region using the CO2SYS \citep{lewis1998program,van2011matlab} subroutine. The CO2SYS subroutine abides by the guidelines set by Ocean Carbon-cycle Intercomparison Project (OCIMP). The sea-surface pH calculated using CO2SYS is referred to as control pH (pH\textsubscript{CTRL}). 

We perform four sensitivity experiments to understand the effect of each of the governing factors (SST, SSS, DIC, and TALK) on the seasonal variability of seas-surface pH. The annual climatological mean is provided to the variable of interest, while other factors are allowed to evolve throughout the year. We refer to the sensitivity cases by pH\textsubscript{X}, where X represents each governing factor. The difference between pH\textsubscript{CTRL} and the pH\textsubscript{X} divulge the effect of the variable or governing factor X.

\subsection{Observational data used for model evaluation}
\label{s:2.3}

We use two observational datasets to evaluate the modeled pH. The description of these datasets are as follows:

\subsubsection{Rama Buoy Data}
\label{s:2.3.1}

The only buoy data available in the BoB region is located at 15\textdegree{} N, 90\textdegree{} E (as shown in Fig.\ref{fig:1}). This buoy measures the ocean surface pH, \textit{p}CO\textsubscript2, salinity, and temperature. The data from this Rama buoy is available from 24 November 2013 to 20 November 2018. Though the buoy is reactivated from January 2020, the data is not available for the public. \cite{joshi2020configuration} provide a table indicating the three deployments of this buoy within the aforementioned period. The method used for calculating the carbonate and physical variables is described meticulously by \cite{sutton2014high}. Unfortunately, pH is the least continuous data measured by the Rama buoy.

\begin{figure}[ht!]
\centering
\includegraphics[height=8cm,width=\textwidth]{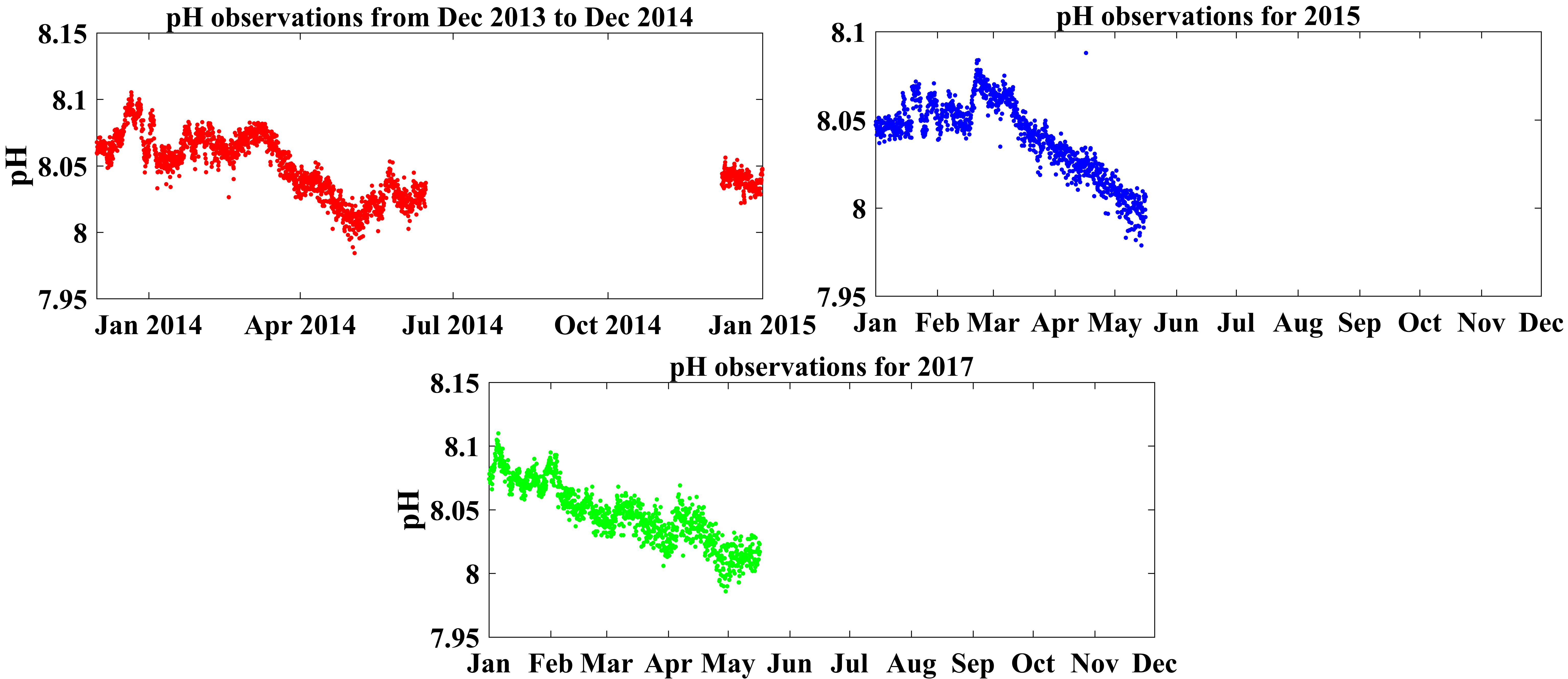}
\caption{The observation pH data for the years 2014, 2015, and 2017 from Rama buoy.}
\label{fig:3}
\end{figure}

Fig.\ref{fig:3} shows the available pH data from Rama buoy. The paucity of real time observation is clearly depicted through Fig.\ref{fig:3}. January to May is a common period between these three years, the pH data for December is available for 2013 and 2014, whereas the pH data for June is only present in 2014. However, in-spite of the limitations recent studies have utilized the RAMA mooring data for validation purpose \citep{joshi2020configuration,sridevi2021role,chakraborty2021seasonal}. To validate our climatology model, we average the available observation data (Fig.\ref{fig:5}).

\subsubsection{Takahashi Data}
\label{s:2.3.2}

Due to the absence of observational data in the BoB region, we choose the monthly pH climatology of \cite{takahashi2014climatological} to evaluate the model performance. This data will be referred to as the Takahashi Data in this manuscript and has been used in previous studies for validating modeled pH \citep{joshi2020configuration,sridevi2021role,chakraborty2021seasonal}. The pH monthly climatology is calculated using the observed \textit{p}CO\textsubscript2 and calculated potential alkalinity (PALK) data. The PALK is calculated using a PALK-salinity relationship, which has shown an inability to capture the effect of the high freshwater influx in the BoB region \citep{takahashi2014climatological}. Hence the pH climatology from Takahashi data showed a significant deviation from the observed pH of the BoB region. The horizontal resolution of 4\textdegree{} $\times$ 5\textdegree{} does not allow the effect of small-scale activities to be adequately captured. Despite the limitations of the Takahashi data, the paucity of observed pH data justifies its use for evaluating the model performance.  

\subsection{Statistics used to aid in model evaluation}
\label{s:2.4}

\begin{table}[H] 
\centering
\begin{tabular}{|P{3cm}| P{9cm} |}
\hline
\textbf{Statistical Parameters} &  \textbf{Description}  \\
\hline
\hline
Correlation Coefficient (r) & Compares the variability between the model and observation data. It ranges from -1 to +1, where the negative value indicates an inverse relationship while the positive values represent a positive relationship.  \\
\hline
Average Error (AE) & \\
Absolute Average Error (AAE) & These parameters represent the bias between observation and model data. \\
Root Mean Square Error (RMSE) & \\
\hline
Reliability Index (RI) & The mean deviation between the model and observed value is given by RI. A value near zero indicates the model to be reliable \citep{stow2009skill}. \\
 \hline
Cost Function (CF) & It is a measure of "goodness of fit." It rates the model in the following manner \citep{dabrowski2014operational} : \\
& CF \ensuremath{<} 1 = excellent ;  1 \ensuremath{\leq} CF \ensuremath{\leq} 2 = good \\
&  2 \ensuremath{\leq} CF \ensuremath{\leq} 3 = average ; CF \ensuremath{>} 3 = poor \\
 \hline 
Percentage Bias (PB) & Similar to CF it is a performance indicator which rates the model in the following manner \citep{dabrowski2014operational} : \\
& PB \ensuremath{<} 10 = excellent ; 10 \ensuremath{\leq} PB \ensuremath{\leq} 20 = very good \\
& 20 \ensuremath{\leq} PB \ensuremath{\leq} 30 = good ; PB \ensuremath{>} 30 = poor \\
\hline
Model Efficiency Factor (MEF) & It determines the predictability of the model concerning mean observed values. MEF quantifies the model efficiency in the following manner \citep{loague1991statistical} :\\
& MEF \ensuremath{\geq} 0.65 = excellent ; 0.5 \ensuremath{\leq} MEF \ensuremath{<} 0.65 = very good \\
& 0.2 \ensuremath{\leq} MEF \ensuremath{<} 0.5 = good ; MEF \ensuremath{<} 0.2 = poor \\
 \hline
\end{tabular}
\caption{Statistical parameters used in this study.}
\label{tab:1}
\end{table}

Table \ref{tab:1} provides a list of statistical indices employed in the present study to evaluate the model performance. The formulation of each of these parameters is provided in Appendix A of \cite{joshi2020configuration}; hence it is not repeated here.

\section{Results and discussion}
\label{s:3}

\subsection{Model Evaluation}
\label{s:3.1}

The present modeled pH performance is evaluated against the two observational data described in sec\ref{s:2.3}. When comparing the model and observation data, we interpolate the model data to the grid resolution of the observation data using the \say{nearest-neighbor interpolation} method. 

 \begin{figure}[ht!]
\centering
\includegraphics[height=8cm,width=11cm]{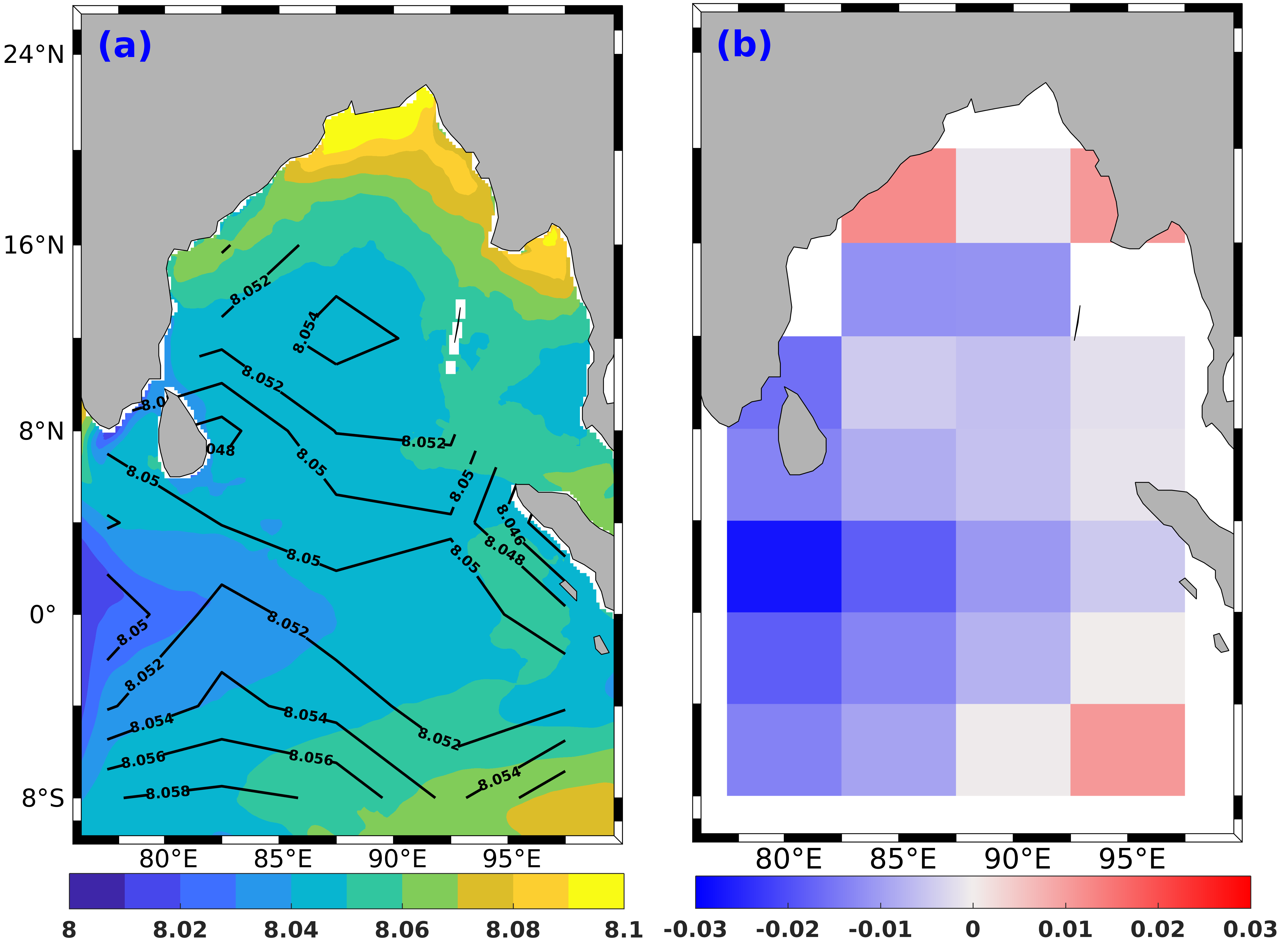}
\caption{In panel (a), the model annual mean pH (color contour) is overlaid with the annual mean pH from the Takahashi data (black lines), and the bias (Model - Observation) is shown in panel (b).}
\label{fig:4}
\end{figure}

Fig.\ref{fig:4}a shows the modeled annual mean pH for the BoB overlaid with the observations from Takahashi data. We observe that the model data shows high spatial variability of the sea-surface pH, while the spatial variability in the Takahashi data is low. The high pH value ($\approx$ 8.08 units) observed from the model in the northern region indicates the alkaline nature of the ocean. The nature of the southern region is comparatively acidic ($\approx$ 8.04 units). The low salinity in the northern region due to the high discharge from rivers and the precipitation could contribute to the high pH values in the north \citep{chakraborty2021seasonal}. As mentioned in sec\ref{s:2.3.2}, the Takahashi data fails to account for the freshwater, which may be attributed to the absence of high pH values from the Takahashi data. The model tends to have a positive bias in the north ($\approx$ 0.01 units), as shown in Fig.\ref{fig:4}b. The pH gradually decreases southwards (Fig.\ref{fig:4}a), this could be due to the high acidic and less discharge from the peninsular rivers. The high saline waters of the Arabian Sea enters the simulation domain decreasing the pH ($\approx$ 8.02 units) in the southwestern region. 

 \begin{figure}[ht!]
\centering
\includegraphics[height=8cm,width=9cm]{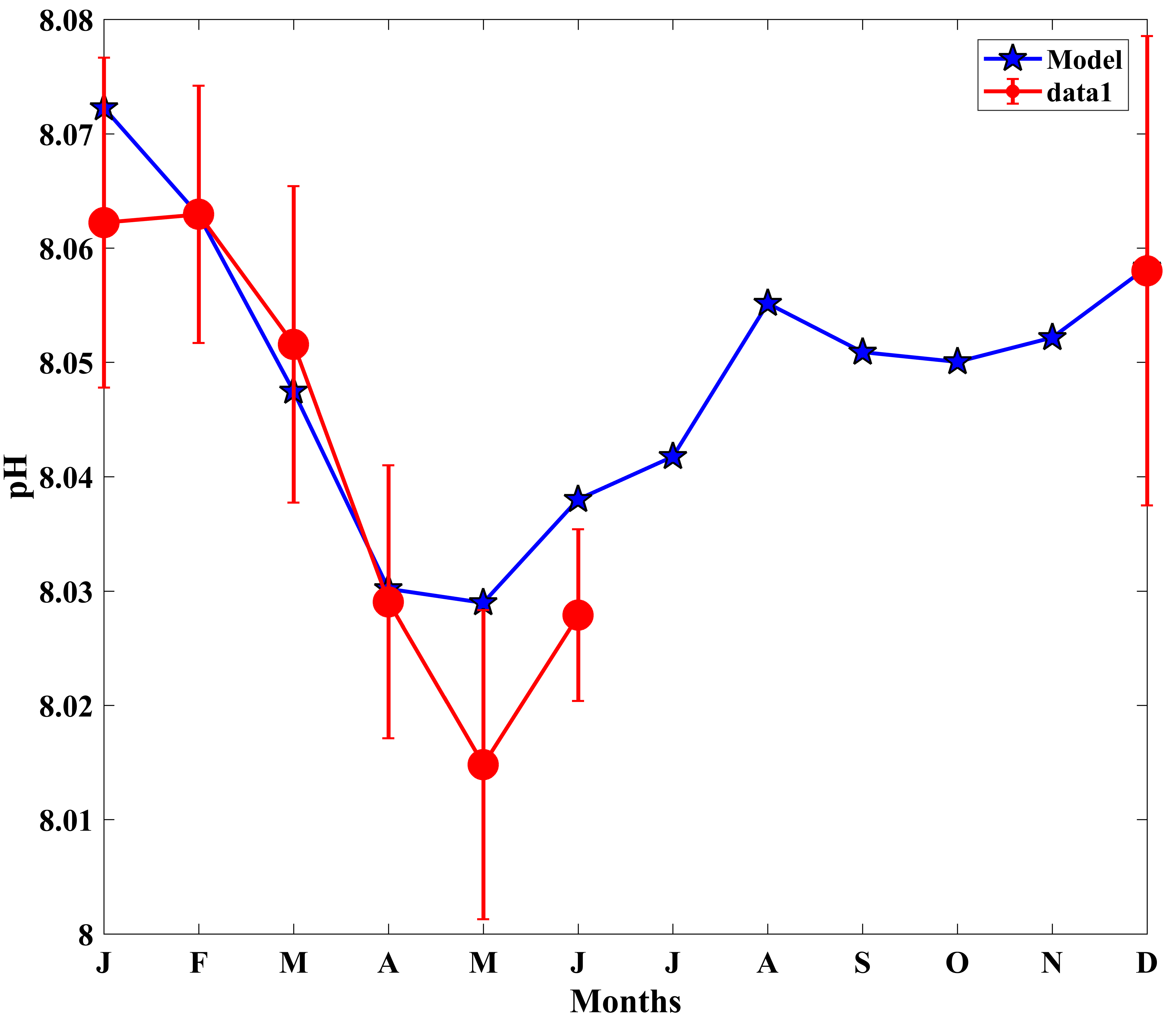}
\caption{Monthly variation in the sea-surface pH from the model and the RAMA buoy.}
\label{fig:5}
\end{figure}

We compare The monthly variability of the modeled sea-surface pH for BoB against the Rama buoy data in Fig.\ref{fig:5}. The paucity of data in the BoB region adheres as a significant hindrance while evaluating the model. Despite the observational data limitations, the model emulates the monthly cycle variability satisfactorily while comparing with buoy data. The model shows a overestimation in May and June, which could be due to the unavailability of recent years data (since the pH is increasing in the central BoB \citep{sridevi2021role}) and relatively higher vertical mixing in the model \citep{joshi2020configuration,joshi2021influence}. The high peak pH value from model pH data noted in the December month could be due to the maximum thickness of the barrier layer \citep{joshi2021influence} inhibits mixing of acidic sub-surface waters with the sea-surface. The FPS distributes the freshwater over the sea-surface of the BoB region, which may affect the solubility and dissociation constants (as they are functions of temperature and salinity). The FPS begins in June, peaking over the post-monsoon season. Thus this freshwater spread may have a significant role in determining the seasonal variability of the pH in BoB. To quantify the model performance, we use different statistical indices as mentioned in sec \ref{s:2.4} (Table \ref{tab:2}).

  \begin{table}[ht!]
     \centering
     \begin{tabular}{|c|c|c|c|c|c|c|c|}
      \hline
      \multicolumn{8}{|c|}{\textbf{Model Vs. RAMA buoy data}} \\
      \hline
      \hline
       AAE & AE & CF & r & RMSE  & PB & RI & MEF\\
      \hline
      \hline
      0.006 & 0.005 & 0.316 & 0.94 & 0.0063 & 0.056 & 1.0002 & 0.87 \\
      \hline
     \end{tabular}
     \caption{Statistical comparison of model-simulated pH with the RAMA buoy data.}
     \label{tab:2}
 \end{table}
 
 A reasonably low bias (0.006) is found when comparing the modeled pH with the Rama buoy data. A high correlation with buoy data indicates that the model well captures the monthly variability. A low reliability index indicate the model to be reliable for modeling the sea-surface pH. The model is rated \say{excellent} by all the other statics indices. Especially high MEF shows the model can replicate the sea-surface pH. The good performance of the model against the RAMA buoy data gives us the confidence to continue further analysis of the sea-surface pH of the BoB region. Though we agree that the data is less but this approach to validate modeled pH is popularly adopted in many past studies \citep{joshi2020configuration,sridevi2021role,chakraborty2021seasonal}.
 
Thus this section highlights the need of more pH observation data in the BoB. We also observe that the SSS may modulate the seasonality of sea-surface pH in the BoB. As the model performance seems fair, we attempt to evaluate and analyze various factors driving the sea-surface pH in the BoB region using this model.
 
 \subsection{Effect of individual drivers on the seasonality of pH}
 \label{s:3.2}
 
  \begin{figure}[ht!]
\centering
\includegraphics[height=8cm,width=\textwidth]{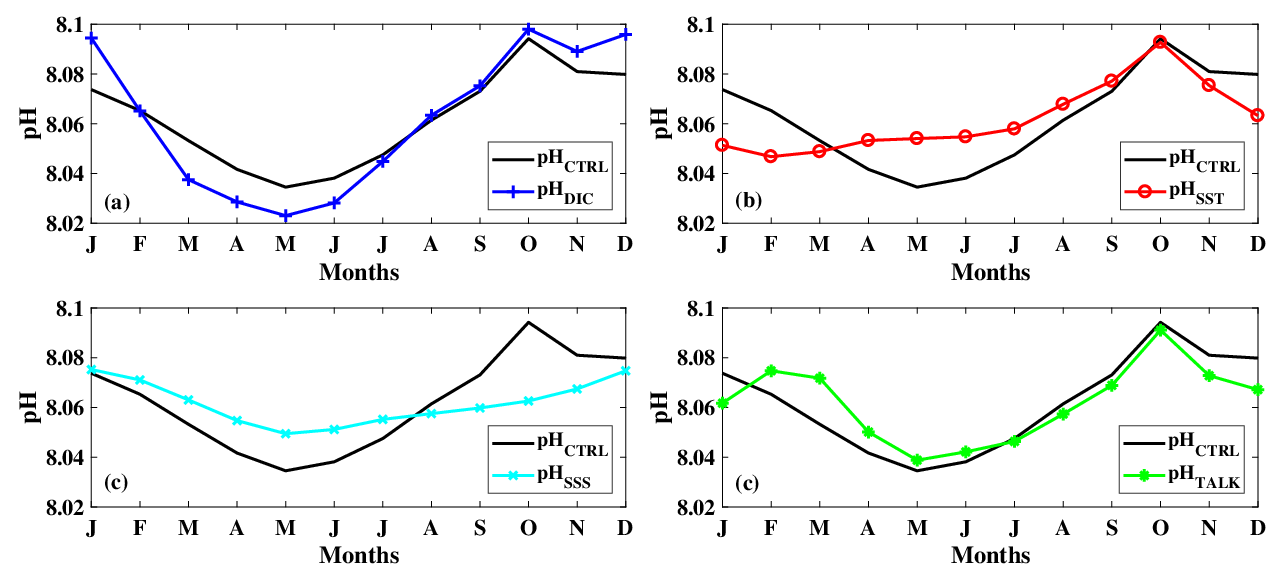}
\caption{Comparison between Control pH (pH\textsubscript{CTRL}) and sensitivity of each drivers (pH\textsubscript{X}). 'X' denotes individual driver (SST, SSS, DIC, and TALK).}
\label{fig:6}
\end{figure}
 
 Fig.\ref{fig:6} presents the effect of individual drivers, viz, SST, DIC, TALK, and SSS on the domain (yellow box, shown in Fig.\ref{fig:1}) averaged pH. Each panel in Fig.\ref{fig:6} shows the effect of these individual drivers on pH, respectively. The pH\textsubscript{CTRL} is shown with a solid black line in each of the Fig.\ref{fig:6} panels. The seasonality in pH\textsubscript{CTRL} seems to be driven by 6 and 12 months signal. We observe a maximum peak of pH\textsubscript{CTRL} in October (8.0942) and a minimum crest in May (8.04). It implies that the sea-surface of BoB is most acidic in May and most alkaline in October. The annual mean pH of the BoB is 8.0620.
 
 In Fig.\ref{fig:6}a, we show the effect of DIC on the pH by replacing it with the annual mean (as described in sec \ref{s:2.2}). The pH\textsubscript{DIC} is significantly higher than the pH\textsubscript{CTRL} during the period between December and February. This high value indicates that the role of DIC is to make the BoB acidic. From March-May, the pH\textsubscript{DIC} is lower than pH\textsubscript{CTRL}, suggesting that the DIC tends to make the surface of BoB alkaline. Interestingly, from June to November, the effect of DIC is negligible; in fact, the primary variation to the ocean acidification of BoB by DIC is found in the first half of the year. The second half of the year shows an almost negligible effect of DIC on pH\textsubscript{CTRL}.  
 
 Fig.\ref{fig:6}b exhibits the effect of temperature (Temp) on the pH. We replace the monthly variation of Temp by annual mean and reproduce pH\textsubscript{SST}. From November to February the pH\textsubscript{SST} is lower than pH\textsubscript{CTRL}, which implies that during this period, the effect of SST is to reduce the acidic nature of BoB. It should be noted that SST tends to drop from November to February (winter months); this may be the reason that during this period, the effect of SST is to make the sea-surface water alkaline. From March to October, the pH\textsubscript{SST} is observed to be higher than pH\textsubscript{CTRL}; hence it reveals that the effect of SST is to make the sea-surface acidic in this period. The higher SST during these months may enhance the acidic nature of the BoB.
 
 The BoB region experiences high freshwater influx, and the effect of salinity (SSS) is revealed in Fig.\ref{fig:6}c. It is clear that for the BoB region, the Sal plays a significant role in modulating the sea-surface pH. From January to August, the pH\textsubscript{SSS} is higher than pH\textsubscript{CTRL}, which suggests that the role of Sal is to increase the acidic levels during this period. Whereas from August-December, the lower values of pH\textsubscript{SSS} with respect to pH\textsubscript{CTRL} reveal that Sal makes BoB alkaline during this period. These results suggest that stratification and the spreading of freshwater plume could possibly heavily modulate the values and dynamics of the sea-surface pH.
 
\begin{figure}[ht!]
\centering
\includegraphics[height=11cm,width=12cm]{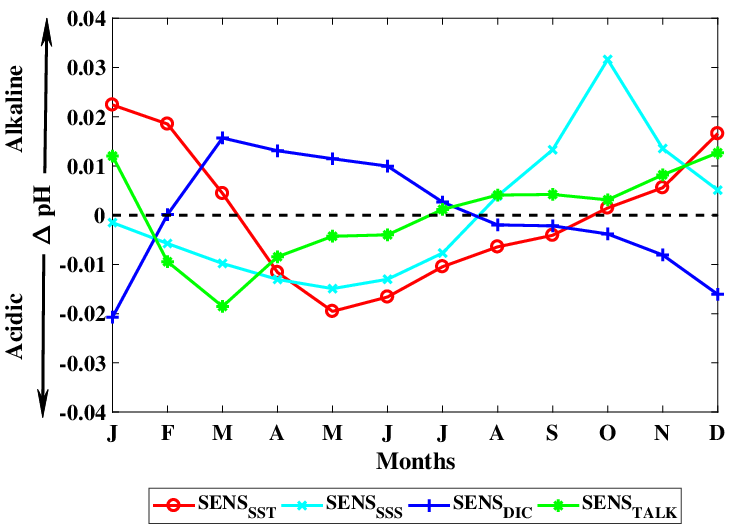}
\caption{Difference between pH\textsubscript{CTRL} and pH\textsubscript{X}, where 'X' represents each driver. SENS\textsubscript{X} defines the sensitivity of the drivers SST, SSS, DIC, and TALK.}
\label{fig:7}
\end{figure}

 Fig.\ref{fig:6}d presents the effect of TALK on the sea-surface pH. We observe that the TALK effect is only significant in February and March. The effect of TALK is almost negligible for most of the climatological year. So it is evident from Fig.\ref{fig:6} that in the BoB region, DIC, SST, and SSS are the major drivers of the seasonality of pH, while TALK has a negligible effect. Further, Fig.\ref{fig:7} summarizes the results of Fig.\ref{fig:6} to quantify the effect of each driver. 
 
Fig.\ref{fig:7} reveals a summary of the effect of each driver on the sea-surface pH of the BoB. Unlike the studies related to the Arabian Sea \citep{sreeush2019variability} and the observational studies from the Pacific ocean, Mediterranean Sea, and North-Atlantic \citep{hagens2016attributing}, the BoB pH is significantly affected by SSS, SST and DIC. Whereas SST and DIC were the major drivers of pH in the Arabian Sea, Pacific Ocean, North Atlantic Ocean, and Mediterranean sea \citep{hagens2016attributing,sreeush2019variability}. From Fig.\ref{fig:7}, we observe that except for February to March period, the effect of TALK is negligible. In February to March period, the effect of DIC and SST is balanced by SSS and TALK. During December and January, the effect of SST is compensated primarily by DIC, whereas both TALK and SSS have negligible effect. The period of August to November is strictly dominated by SSS, making the sea-surface alkaline. The peak in October coincides with the maximum freshwater plume spread \citep{jana2015impact,jana2018sensitivity,joshi2021influence}, which indicates that the freshwater which lowers the SSS increases the pH of the BoB region. The acidic peak in May can be attributed to the highest SSS in the BoB region and lowest freshwater plume spread \citep{jana2015impact,jana2018sensitivity,joshi2021influence}. This domination of SSS in controlling pH in the BoB region makes this region unique from the rest of the world ocean.

\begin{figure}[ht!]
\centering
\includegraphics[height=10cm,width=\textwidth]{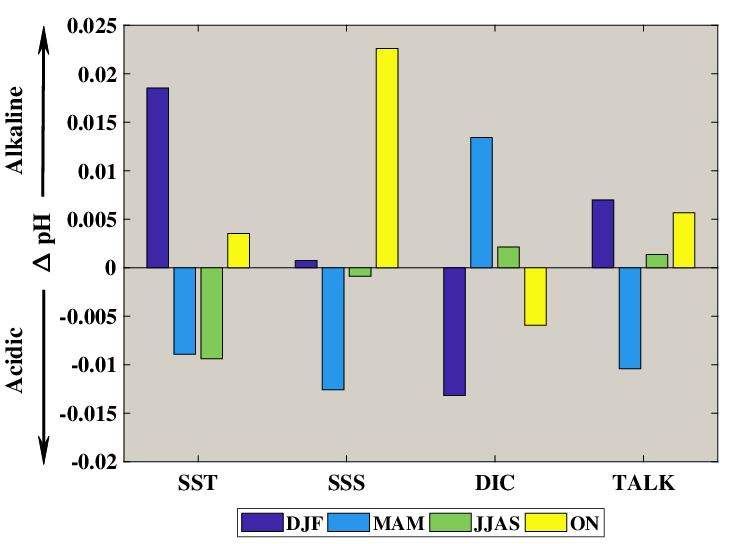}
\caption{Seasonal effect of the drivers of pH.}
\label{fig:8}
\end{figure}

In Fig.\ref{fig:8}, we demonstrate the seasonal effect of each of the governing factors on the sea-surface pH of the BoB region. The climatological year is divided based on the monsoon season. The December to February (DJF) is the winter monsoon season, the June to September (JJAS) is the southwest monsoon season, while the phases March to May (MAM) and October-November (ON) are pre-monsoon and post-monsoon seasons, respectively. In the winter monsoon season (DJF), the SST tends to increase the pH, while the DIC averse the rise in pH, but the minor effect of TALK further increments the pH. The effect of SSS is almost zero in the winter monsoon season. The pre-monsoon season (MAM) experiences the combined effect of all the drivers. The DIC tends to increase the pH during MAM, but the combined effect of SSS, SST, and TALK reduces the pH. Thus the pre-monsoon season is the most acidic season among all other seasons.  

In the southwest monsoon season (JJAS), pH is dominated by SST. The SST tends to reduce the pH in the southwest monsoon season. The effect of SSS, DIC, and TALK is minimal during the southwest monsoon season. The SSS dominates the rise in pH during the post-monsoon season. This may be attributed to the freshwater plume spread during this period \citep{jana2015impact,jana2018sensitivity,joshi2021influence}. The DIC effect is countered by the TALK in the post-monsoon season, while the SST has a negligible effect on pH during this season. 

Hence, our analysis shows that SSS is a major driver of pH along with SST and DIC in the BoB region. The role of SSS in modulating the pH may be due to the high freshwater flux, strong stratification, and thick barrier layer, inhibiting vertical mixing. The seasonality in SSS and SST also affects the sea-surface pH of BoB. As SSS has a significant role in modulating the seasonality of pH in the BoB region, further analysis on the effect of stratification and barrier layer over the pH becomes highly important.

\subsection{Role of stratification on seasonality of pH}
\label{s:3.3}

Due to the low-density freshwater influx from rivers and precipitation, a shallow layer of freshwater forms in the sub-surface waters, leading to the strong stratification in the BoB region  \citep{shetye1996hydrography,vinayachandran2002observations,rao2003seasonal}. The strong stratification leads to a barrier layer formation. \cite{shetye1993movement,shetye1996hydrography} reported observing a low saline strip of freshwater along the western coast in July-August 1989 and December 1991, respectively. The formation and spreading of the freshwater plume in the summer monsoon are meticulously described through observation \citep{vinayachandran2007hydrographic}. \cite{jana2015impact,jana2018sensitivity,sandeep2019riverine} explain the role of winds, circulation, and rivers from a modeling perspective, in the formation and spreading of the freshwater plume. The freshwater spread over the surface of BoB, and the barrier layer thickness (BLT) thus formed restricts vertical mixing. This inhibition of vertical mixing does not allow the sub-surface nutrients and chemicals from reaching the surface, which makes the BoB region a low-productive zone \citep{prasanna2002bay,gauns2005comparative}. Our previous work \citep{joshi2021influence} shows the extent by which BLT and the freshwater plume influences the sea-surface \textit{p}CO\textsubscript2. The method to calculate the BLT is same for this study, hence we encourage readers to refer to our previous work for detailed explanation of BLT calculation and its influence on the sea-surface \textit{p}CO\textsubscript2.

\begin{figure}[ht!]
\centering
\includegraphics[height=10cm,width=10cm]{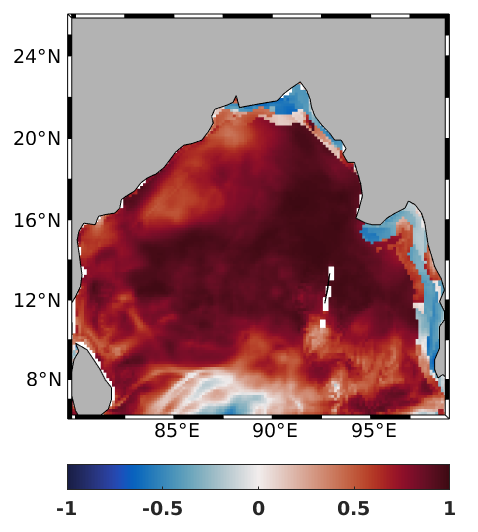}
\caption{Spatial correlation between BLT and sea-surface pH.}
\label{fig:9}
\end{figure}

To examine the relationship between the BLT and the sea-surface pH in the BoB region, we make a spatial correlation between the BLT and pH, as shown in Fig.\ref{fig:9}. The correlation is used as a proxy to highlight the relation between BLT and pH. As we observe from Fig.\ref{fig:9}, the open oceans of the BoB region shows a high positive correlation between pH and BLT, suggesting that the thicker the barrier layer higher the sea-surface pH. This may suggest that the thick barrier layer impedes the mixing of sub-surface acidic waters with the surface waters of BoB, resulting in higher pH at the sea-surface. The high positive correlation also suggests that the months or seasons having low BLT would result in acidic sea-surface waters. 

\begin{figure}[ht!]
\centering
\includegraphics[height=14cm,width=\textwidth]{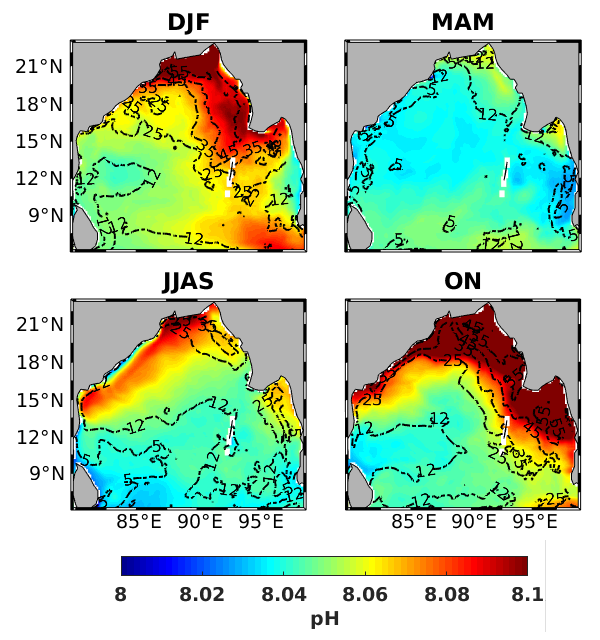}
\caption{The seasonal spatial plot showing pH (contours) overlaid with BLT in black broken lines.}
\label{fig:10}
\end{figure}

In Fig.\ref{fig:10}, we emulate the spatial seasonal pH overlaid with the seasonal BLT. We observe that indeed the areas having thicker barrier layer seem to have higher pH values. The barrier layer is known to be thickest during the winter monsoon period (DJF), and the area having thicker barrier layers is spread the most during this season. Hence the higher pH value area is seen to be more widely spread during the winter monsoon season. The pre-monsoon season (MAM) is observed to have the shallowest barrier layer, and hence the interaction between sub-surface and surface waters must be predominant. Thus the pre-monsoon season seems to be the season having the most acidic sea-surface. In Fig.\ref{fig:8}, we see that the pH in the pre-monsoon season has a significant contribution from all the governing factors, the interaction between sub-surface and surface waters may contribute to the equal roles of all drivers.

\begin{figure}[ht!]
\centering
\includegraphics[height=12cm,width=12cm]{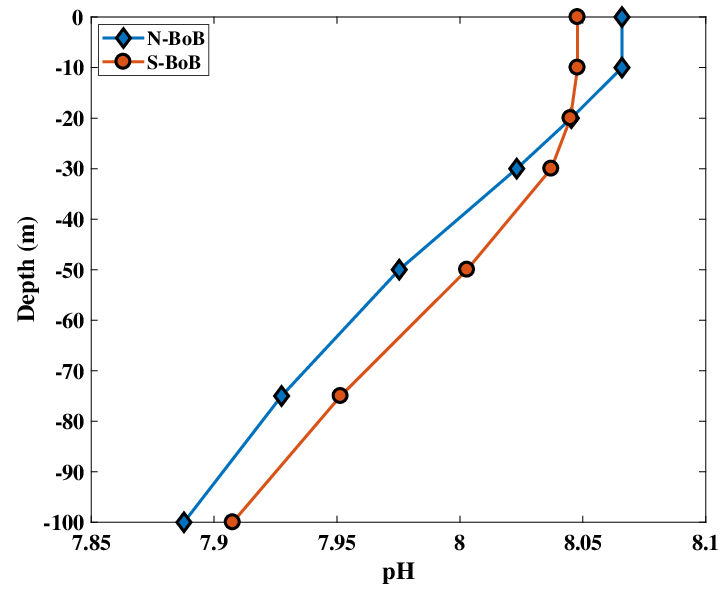}
\caption{Comparison of the depth-wise pH profile between the northern (N-BoB) and southern (S-BoB) BoB.}
\label{fig:11}
\end{figure}

The southwest monsoon season (JJAS) in Fig.\ref{fig:10} exhibits a strip of low acidic waters slanting from the north BoB along the western coast. The beginning of freshwater plume spread in the southwest monsoon season seems to draw the low-saline waters through the west coast \citep{shetye1993movement}, which increases the pH in this region. The post-monsoon season (ON) marks the largest freshwater plume spread and the lowest SSS, resulting in the highest pH, especially in the northern region. It is observed from Fig.\ref{fig:10} that the northern region of BoB (as shown in Fig.\ref{fig:1}) has higher pH values except for the pre-monsoon season. The freshwater influx from high discharge rivers like Ganga, Brahmaputra, and Irrawaddy, along with the high precipitation, maybe why the high pH and high BLT values in the north. The southern sea-surface is comparatively acidic in nature. Especially the southwestern region seems to be relatively acidic throughout the climatological year. To further visualize the effect of stratification, we plot the depth-wise annual mean pH (up to 100m) (Fig.\ref{fig:11}) for the N-BoB and S-BoB demarcated by the broken black line in Fig.\ref{fig:1}.

We observe in Fig.\ref{fig:11} that the pH in the N-BoB is lower than the S-BoB below 20m depth. On the surface, the N-BoB has a pH higher than the S-BoB. We know that the N-BoB is more influenced by the freshwater influx due to the BLT and consequently stronger stratification; hence the acidic lower pH waters are restricted at around 20 to 22m depth. The pH in N-BoB is constant till 10m depth, whereas the S-BoB pH is constant almost till 20m depth, which indicates that the inhibition of vertical mixing in N-BoB is lower than the S-BoB. Hence the effect of stratification in the N-BoB plays a vital role in making the N-BoB sea-surface alkaline compared to the S-BoB.

\section{Conclusion}
\label{s:4}

In the present study, we attempt to explore the climatological ocean acidification state of the BoB region. We use the outputs of a coupled physical-biogeochemical model to reconstruct the pH for the BoB region. The paucity of data from the chemical parameters has been a challenge to establish the model worldwide. The high freshwater influx in the BoB region makes this region even more challenging as mathematical relations of chemical parameters based on SSS do not stand valid in this region. Nevertheless, considering these challenges, we try to validate our modeled pH with the limited data available. The modeled pH performs reasonably well when compared with RAMA buoy and Takahashi data. The correlation of 0.94 with RAMA buoy implies that the model well captures the monthly variability. The biases have been reasonably low when comparing with both the observational data. The comparison also highlights the necessity of more real-time chemical parameter observations in the BoB region.  

Further, we analyze the role of each governing factor, viz, SST, SSS, DIC, and TALK, in modulating the seasonality of pH. The SSS has emerged as one of the major drivers of pH along with SST and DIC. The first six months of the climatological year have a combined effect of SST, DIC, and SSS on pH, but SSS significantly dominates the last six months. The SST reduces the sea-surface acidity in the winter monsoon season and post-monsoon season, but it increases the acidity in pre-monsoon and southwest monsoon seasons. The winter monsoons experience the maximum effect of SST on pH, and the post-monsoon season has a minimum SST effect on pH. SSS increases the pH significantly in the post-monsoon season. The SSS reduces the pH in the pre-monsoon season. The effect of SSS is minor during the southwest monsoon and winter monsoon seasons. DIC increases acidity in the sea-surface during the winter monsoon season but reduces acidity during the pre-monsoon season. The effect of TALK is lowest among all the governing factors.

As SSS is a major contributor in modulating the seasonality of pH, the role of stratification on sea-surface ocean acidification becomes crucial in the BoB region. The sea-surface pH shows a strong positive correlation with the BLT, which indicates that thicker barrier layer areas have higher sea-surface pH. The N-BoB is found to be more alkaline than the S-BoB. The spreading of the freshwater plume that reduces the SSS may be why the higher pH values in the N-BoB. Below 20 m depth, the N-BoB has more acidic waters than the S-BoB. The stronger stratification and thick barrier layer inhibit these high acidic sub-surface waters from reaching the surface resulting in a more alkaline sea-surface in N-BoB than the S-BoB.

Our study is one of the first climatological modeling approaches for the pH in the BoB region. The study becomes more important as it provides a significant idea of the governing factors of pH in the open ocean part of BoB. With global warming, these roles may change in the future. The relationship between BLT and sea-surface pH gives a different perspective which may be used to artificially (using a neural network approach) generate pH data. However, interannual modeling has to be done, as cyclones and higher local winds may change the role of stratification on pH. In the next part of this study, we will evaluate the effect of other mechanisms like biological production, solubility, and carbon dioxide flux on the sea-surface pH.

\section*{Acknowledgment}

The present work is carried out at the Indian Institute of Technology, Kharagpur. The authors extend gratitude towards the institute for providing support for this work. We also thank Prof. Arun Chakraborty for introducing us to the Regional Ocean Model (ROMS). The first author acknowledges the Ministry of Education, India, for providing the scholarship assistance. 
The RAMA buoy pH data is downloaded from \url{https://www.nodc.noaa.gov/ocads/oceans/Moorings/BOBOA.html}. The Takahashi data is downloaded from \url{https://www.nodc.noaa.gov/ocads/oceans/ndp\_094/ndp094.html}. There is no conflict of interest between the authors to declare.






\bibliographystyle{elsarticle-harv} \biboptions{authoryear}
\bibliography{sample.bib}







\end{document}